
\input phyzzx
\hoffset=0.2 true in
\voffset=0.2 true in
\def\lapp{{\ \lower 0.6ex \hbox{$\buildrel<\over\sim$}\ }}
\def\gapp{{\ \lower 0.6ex \hbox{$\buildrel>\over\sim$}\ }}
\rightline{SNUTP 94--58}
\rightline{YUMS 94--18}
\rightline{(June 1994)}
\bigskip\bigskip

\nopagenumbers
\title{\bf  Fermi motion parameter $p_{_F}$ of $B$ meson  }
\title{\bf from relativistic quark model }
\bigskip\bigskip\bigskip
\centerline{Daesung Hwang$^a$, C.S.~~Kim$^a$\foot{E-address:
kim@cskim.yonsei.ac.kr} and Wuk Namgung$^b$}
\bigskip\bigskip
\centerline{$a$ Department of Physics, Yonsei University, Seoul 120-749, KOREA
}
\centerline{$b$ Department of Physics, Dongguk University, Seoul 100-715, KOREA
}
\bigskip\bigskip\bigskip

\centerline{\bf ABSTRACT}
\medskip

The Fermi  motion parameter $p_{_F}$ is the most important parameter of
ACCMM model, and the value $p_{_F} \sim 0.3$ GeV has been used without clear
theoretical or experimental evidence. So, we attempted to extract the possible
value for $p_{_F}$ theoretically in the relativistic quark model using  quantum
mechanical variational method. We obtained $p_{_F} \sim 0.5$ GeV, which is
somewhat larger than 0.3, and we also derived the eigenvalue of $E_B \simeq
5.5$ GeV, which is is reasonable agreement with $m_B=5.28$ GeV.

\vfill\eject

\pagenumbers
\pagenumber=2
\baselineskip = 20 pt

In the minimal standard model CP violation is possible through the CKM mixing
matrix of three families, and it is important to know whether the element
$V_{ub}$ is
non-zero or not accurately. Its knowledge is also necessary to check whether
the
unitarity
triangle  is closed or not [1].
However its experimental value is very poorly known until now and its better
experimental information is urgently required. At present, the only
experimental method
to study $V_{ub}$ has been through the end-point leptonic spectrum of the
B-meson
semileptonic decays of CLEO [2] and ARGUS [3], and their data
indicate that $V_{ub}$ is non-zero.
Recntly it has also been  suggested [4] that the measurement of hadronic
invariant
mass distribution in the inclusive $B \rightarrow X_{c(u)} l \nu$ decays can be
useful in extracting $|V_{ub}|$ with better theoretical understandings.
In future asymmetric $B$ factories with vertex detector, this will offer
an alternative way to select $b \rightarrow u$ transitions that is much more
efficient than selecting the upper end of the lepton energy spectrum.

The simplest model for the semileptonic B-decay is the spectator model which
considers the decaying $b$-quark in the B-meson as a free particle.
The spectator model
is usually used with the inclusion of perturbative QCD radiative corrections.
The decay
width of the process $B\rightarrow X_ql\nu$ is given by
$$
{\Gamma}_B (B\rightarrow X_ql\nu )\simeq {\Gamma}_b (b\rightarrow ql\nu )=
\vert V_{bq}{\vert}^2({{G_F^2m_b^5}\over {192{\pi}^3}})f({{m_q}\over {m_b}})
[1-{{2}\over {3}}{{{\alpha}_s}\over {\pi}}g({{m_q}\over {m_b}})]~~,
\eqno(1)
$$
where $m_q$ is the mass of the $q$-quark decayed from $b$-quark.
The decay width of the spectator model depends on $m_b^5$, therefore small
difference of $m_b$ would change the decay width significantly.

Altarelli et al. [5] proposed their ACCMM model for the inclusive B-meson
semileptonic decays. This model incorporates the bound state
effect by treating the $b$-quark as a vitual state particle, thus giving
momentum dependence to the $b$-quark mass. The virtual state $b$-quark mass
$W$ is given by
$$
W^2({\bf p})=m_B^2+m_{sp}^2-2m_B{\sqrt{{\bf p}^2+m_{sp}^2}}
\eqno(2)
$$
in the B-meson rest frame, where $m_{sp}$ is the spectator quark mass,
$m_B$ the B-meson mass, and {\bf p} is the momentum of the $b$-quark.

For the momentum distribution of the virtual $b$-quark, Altarelli et al.
considered the Fermi motion inside the B-meson as the Gaussian
distribution
$$
\phi ({\bf p})={{4}\over {{\sqrt{\pi}}p_{_F}^3}}e^{-{\bf p}^2/p_{_F}^2}
\eqno(3)
$$
with a free parameter $p_{_F}$ of Gaussian width.
And the decay width is given by integrating the width ${\Gamma}_b$ in
(1) with the weight $\phi ({\bf p})$. Then the leptonic spectrum
of the B-meson semileptonic decay is given by
$$
{{d{\Gamma}_B}\over {dE_l}}(p_{_F}, m_{sp}, m_q, m_B)=
{\int}_0^{p_{max}}dp\ p^2\phi ({\bf p})\
{{d{\Gamma}_b}\over{dE_l}}(m_b=W, m_q)~~,
\eqno(4)
$$
where $p_{max}$ is the maximum kinematically allowed value of $p=|{\bf p}|$.
The ACCMM model, therefore,
introduces a new parameter $p_{_F}$ for the momentum measure of the
Gaussian distribution inside B-meson instead of the $b$-quark mass of
the spectator model.
In this way the ACCMM model incorporates the bound state effects and reduces
the strong dependence on $b$-quark mass in the decay width of the
spectator model.

The Fermi motion parameter $p_{_F}$ is the most essential parameter of
the ACCMM
model as explained above. However, the experimental determinations of its value
from the leptonic spectrum have been very ambiguous until now because the
various effects from the input parameters and the perturbative QCD corrections
are intermixed in the spectrum. The value $p_{_F} \sim 0.3$ has been widely
used in experimental analyses without theoretical or experimental clean
justification, even though there has been recently an assertion
that the BSUV model [6] is approximately equal to ACCMM model at
$p_{_F} \simeq 0.3$.
Therefore, it is strongly required to determine the value of $p_{_F}$ more
firmly when we
think of the importance of its role in experimental analyses. The better
determination of $p_{_F}$ is also interesting theoretically since it has
the physical correspondence related to the Fermi motion inside B-meson.
In this context we are
going to theoretically determine the value of $p_{_F}$ in the relativistic
quark model using quantum mechanical variational method.

We consider the Gaussian probability distribution function $\phi ({\bf p})$
in (3) as
the absolute square of the momentum space wave function $\chi ({\bf p})$ of
the bound state B-meson, i.e.,
$$\eqalignno{
\phi ({\bf p}) &= 4\pi \vert \chi ({\bf p}){\vert}^2~~, &(5) \cr
\chi ({\bf p}) &= {{1}\over {({\sqrt{\pi}}p_{_F})^{3/2}}}e^{-{\bf
p}^2/2p_{_F}^2}~~.
&(6)}
$$
The Fourier transform of $\chi ({\bf p})$ gives the coordinate space wave
function $\psi ({\bf r})$, which is also Gaussian,
$$
\psi ({\bf r})=({{p_{_F}}\over {\sqrt{\pi}}})^{3/2}e^{-r^2p_{_F}^2/2}~~.
\eqno(7)
$$
Then we can approach the determination of $p_{_F}$ in the framework of
quantum mechanics.
We apply the variational method with the Hamiltonian operator
$$
H={\sqrt{{\bf p}^2+m_{sp}^2}}+{\sqrt{{\bf p}^2+m_b^2}}+V(r)
\eqno(8)
$$
and the trial wave function
$$
\psi ({\bf r})=({{\mu}\over {\sqrt{\pi}}})^{3/2}e^{-{\mu}^2r^2/2}~~,
\eqno(9)
$$
where $\mu$ is the variational parameter.
The ground state is given by
minimizing the expectation value of $H$,
$$\eqalignno{
&\langle H\rangle =\langle\psi\vert H\vert\psi\rangle =E(\mu )~~,  &(10) \cr
&{{d}\over {d\mu }}E(\mu )=0\ \ {\rm{at}}\ \ \mu ={\bar{\mu}}~~,  &(11)}
$$
and then ${\bar{\mu}} = p_{_F}$ and $\bar E \equiv E({\bar{\mu}})$
approximates $m_B$.
The value of $\mu$ or $p_{_F}$ corresponds to the
measure of the radius of the two body bound state as can be seen from
$$
\langle r\rangle ={{2}\over{\sqrt{\pi}}}{{1}\over {\mu}}\quad ,
\quad  {\rm and} \quad
\langle r^2{\rangle}^{{1}\over {2}} ={{3}\over {2}}{{1}\over {\mu}}~~.
\eqno(12)
$$

In (8), for simplicity, we take the Cornell potential
which is composed of the coulomb and linear potentials,
$$
V(r)=-{{{\alpha}_c}\over {r}}+Kr~~.
\eqno(13)
$$
For the values of the parameters ${\alpha}_c\ (\equiv {{4}\over
{3}}{\alpha}_s)$, K, and the $b$-quark mass $m_b$, we use the values given by
Hagiwara et al. [7],
$$
{\alpha}_c=0.47\ ({\alpha}_s=0.35),\ \ K=0.19\ GeV^2,\ \ m_b=4.75\ GeV,
\eqno(14)
$$
which have been determined by the best fit of the $(c{\bar{c}})$ and
$(b{\bar{b}})$ bound states.
We will also consider the following value of ${\alpha}_c$ for comparison
in our analysis
$$
{\alpha}_c=0.32\ ({\alpha}_s=0.24) \quad ,
\eqno(15)
$$
which corresponds to $\alpha_s(Q^2 = m_B^2)$.

Before applying our variational method with the Gaussian trial wave function
to the B-meson system, let us check the method by considering the $\Upsilon
(b{\bar{b}})$ system.
The Hamiltonian of the $\Upsilon (b{\bar{b}})$ system can be approximated
by the  non-relativistic Hamiltonian
$$
H\simeq 2m_b+{{{\bf p}^2}\over {m_b}}+V(r)~~.
\eqno(16)
$$
With the parameters in (14) or (15),  our variational method with
the Gaussian trial wave function gives $p_{_F}={\bar{\mu}}=1.1\ GeV$ and
${\bar{E}}=E({\bar{\mu}})=9.49\ GeV$. Here $p_{_F}=1.1\ GeV$ corresponds to the
radius $R(\Upsilon )=0.2\ fm$, and ${\bar{E}(\Upsilon)}=9.49\ GeV$ is within
$0.3\ \%$  error compared with the experimental value $E_{\rm{ exp}}=
m_{\Upsilon}=9.46\ GeV$. Therefore, the
variational method with the non-relativistic Hamiltonian gives the fairly
accurate results for the $\Upsilon$ ground state.

However, since the $u$- or $d$- quark in the B-meson is very light, the
non-relativistic description can not be applied to the B-meson system.
For example, when we apply the variational method with the non-relativistic
Hamiltonian to the B-meson, we get the following results
$$\eqalignno{
p_{_F}=0.29\ GeV,\ \ {\bar{E}} &=5.92\ GeV\ \ {\rm{for}}\ \ {\alpha}_s=0.35,
&(17) \cr
p_{_F}=0.29\ GeV,\ \ {\bar{E}} &=5.97\ GeV\ \ {\rm{for}}\ \ {\alpha}_s=0.24.
&(18)}
$$
The above masses $\bar E$ are much larger compared to the experimental value
$m_B=5.28\ GeV$, and moreover the expectation values of the higher
terms in the non-relativistic perturbative expansion are bigger than those of
the lower terms. Therefore, we can not apply the variational method with the
non-relativistic Hamiltonian to the B-meson system.

We use the following Hamiltonian for the B-meson system in our analysis by
treating $u$- or $d$-quark relativistically,
$$
H=M+{{{\bf p}^2}\over {2M}}+{\sqrt{{\bf p}^2+m^2}}+V(r)~~,
\eqno(19)
$$
where $M=m_b$ and $m=m_{sp}$.

In our variational method the trial wave
function is Gaussian both in the coordinate space and in the momentum space,
so the expectation value of $H$ can be calculated in either space,
$$
\langle H\rangle =\langle\psi({\bf r})\vert H\vert\psi({\bf r})\rangle
=\langle\chi({\bf p})\vert H\vert\chi({\bf p})\rangle~~ .
\eqno(20)
$$
Also, the Gaussian function is a smooth function and its derivative of any
order is square integrable, thus any power of the Laplacian operator
${\nabla}^2$ is a hermitian operator at least under Gaussian functions.
Therefore, analyzing the Hamiltonian (19) with the variational method
can be considered as reasonable even though solving the eigenvalue equation
of the differential operator  (19) may
be confronted with the mathematical difficulties because of the square root
operator in (19).

With the Gaussian trial wave function (6) or (9), the expectation value of $H$
can
easily be calculated besides  the square root operator,
$$
\eqalignno{\langle {\bf p}\,^2\rangle &= \langle \psi ({\bf r}\,) |
{\bf p}\,^2| \psi ({\bf r}\,) \rangle = \langle \chi ({\bf p}\,) | {\bf p}\,^2|
\chi ({\bf p}\,) \rangle = {3 \over 2} \mu^2~~, &(21) \cr
\langle V(r) \rangle &= \langle \psi ({\bf r}) | -{\alpha_c \over r} + Kr
|\psi ({\bf r}) \rangle = {2 \over \sqrt\pi} (-\alpha_c\mu + {K / \mu} ) ~~.
&(22)}
$$
Now let us consider the expectation value of the square root operator  in
the momentum space
$$
\eqalignno{\langle \sqrt{{\bf p}\,^2+ m^2} \rangle &= \langle \chi ({\bf p}\,)
| \sqrt{{\bf p}\,^2+ m^2} | \chi ({\bf p}\,) \rangle \cr
&= \Bigl({\mu \over \sqrt\pi}\Bigr)^3 \int_0^\infty
e^{-{p^2 / \mu^2}} \sqrt{{\bf p}\,^2+ m^2}\; d^3p \cr
&= {4\mu \over \sqrt\pi} \int_0^\infty e^{-x^2} \sqrt{x^2 + (m/\mu)^2} \;
x^2dx ~~. &(23)}
$$
The integral (23) can be given as a series expansion by the following
procedure. First, define
$$
\eqalignno{I(s)\, &\equiv \int_0^\infty \sqrt{x^2 + s} \; x^2 e^{-x^2} dx
= s^2 \int_0^\infty \sqrt{t^2 + 1} \; t^2 e^{-st^2} dt~~, &(24) \cr
I_0(s) &\equiv \int_0^\infty \sqrt{x^2 + s} \; e^{-x^2} dx
= s \int_0^\infty \sqrt{t^2 + 1} \; e^{-st^2} dt~~. &(25)}
$$
Next,  from (24) and (25), we find the following differential relations
$$
\eqalignno{
&{d \over ds} \Bigl({I_0 \over s}\Bigr)  = - {1 \over s^2} I ~~, &(26)\cr
&{d I \over d s}  = - {1 \over 2} I_0 + I ~~. &(27)}
$$
Combining (26) and (27), we get a second order differential equation for
$I(s)$,
$$
s I''(s) - (1+s) I'(s) + {1 \over 2} I(s) = 0~~. \eqno{(28)}
$$
The series solution to eq. (28) is given as
\vfill\eject
$$\eqalignno{
I(s)\;\; &= c_1 I_1 (s) + c_2 I_2 (s)~~, \cr
I_1 (s) \,&= s^2 F(s; {3 \over 2}, 3) = s^2 \Bigl\{ 1 + {1 \over 2} s +
{5 \over 32} s^2 + {7 \over 192} s^3 + {7 \over 1024} s^4 + \cdots\Bigr\}~~,
&(29) \cr
I_2 (s) &= I_1 (s) \int {s e^s \over [I_1 (s)]^2} ds \cr
&= - {1 \over 16} s^2 \ln s \Bigl( 1 + {1 \over 2} s + {5 \over 32} s^2 +
\cdots \Bigr) 
-{1 \over 2} \Bigl( 1 + {1 \over 2} s + {5 \over 32} s^2 +
{7 \over 192} s^3 + {7 \over 1536} s^4 + \cdots \Bigr)~,   }
$$
where $F(s; {3\over 2}, 3)$ is the confluent hypergeometric function which
is convergent for any finite $s$, and the integral constants $c_1\simeq-0.095$,
$c_2=-1$.
See  Appendix for the derivation of these numerical values for $c_i$.

Finally, collecting (21), (22) and (23), the expectation value of $H$ is
written as
$$
\eqalign{\langle H \rangle &= M + {1\over 2M} \Bigl({3\over 2} \mu^2
\Bigr) + {2 \over \sqrt\pi} ( -\alpha_c \mu + K/\mu ) \cr
&\quad + {2\mu \over \sqrt\pi} \biggl[ 1 + {1\over 2} (m/\mu)^2 +
\Bigl({5\over 32} - 2c_1 \Bigr) (m/\mu)^4 + {1\over 4} (m/\mu)^4 \ln(m/\mu)
\biggr] ~~,}
\eqno{(30)}
$$
up to $(m/\mu)^4$.

With  the input value of $m =m_{sp} = 0.15$ GeV, we minimize
$\langle H \rangle$ of (30),  and we obtain
$$
\eqalign{p_{_F}=\bar \mu &= 0.54 \ GeV, \qquad m_B=\bar E = 5.54 \ GeV
\qquad {\rm for}\;
\alpha_s=0.35~~, \cr
\bar \mu &= 0.49 \ GeV, \ \ \qquad\qquad\bar E = 5.63 \ GeV \ \qquad {\rm
for}\;
\alpha_s=0.24~~.}
\eqno{(31)}
$$
For comparison, we calculated $\langle H \rangle$ for the case of $m=0$ in
which the integral of the square root operator is exact,  and we get
$$
\eqalign{\bar \mu = 0.53 \ GeV, \qquad\bar E = 5.52 \ GeV \qquad {\rm for}\;
\alpha_s=0.35~~, \cr
\bar \mu = 0.48 \ GeV, \qquad\bar E = 5.60 \ GeV \qquad {\rm for}\;
\alpha_s=0.24~~.}
\eqno{(32)}
$$
The calculated values of the $B$-meson mass, $\bar E$,  are much larger than
the
measured  value of 5.28.
The large values for the mass are   originated  partly because the Hamiltonian
(30)  does not take care of the correct spin dependences  for $B$ and $B^*$.
The difference between the pseudoscalar meson and the vector meson is  given
arise to by the
chromomagnetic hyperfine splitting, which is 
$$
V_s = {2 \over 3Mm} \; \vec s_1 \cdot \vec s_2 \,\nabla^2
(- {\alpha_c \over r})~~.
\eqno{(33)}
$$
Then the expectation value of $V_s$ is given as
$$
\eqalignno{\langle V_s \rangle &= - {2 \over \sqrt\pi} \,
{\alpha_c \mu^3 \over Mm} \qquad{\rm for} \quad B~~, &(34-1) \cr
&= {2 \over 3\sqrt\pi} \, {\alpha_c \mu^3 \over Mm} \qquad\; {\rm for}
\quad B^*~~,&(34-2) }
$$
and we treat $\langle V_s \rangle$ only as a perturbation.
And we get for $B$ meson
$$
\eqalign{p_{_F} = 0.54 \ GeV, \qquad \bar E_B = 5.42 \ GeV \qquad
&{\rm for}\; \alpha_s=0.35~~,  \cr
p_{_F} = 0.49 \ GeV, \qquad \bar E_B = 5.56 \ GeV\qquad
&{\rm for}\; \alpha_s=0.24~~. }
\eqno(35)
$$
The perturbative result for $B^*$ is
$$
\eqalign{p_{_F} = 0.54 \ GeV, \qquad \bar E_{B^*} = 5.58 \ GeV \qquad
&{\rm for}\; \alpha_s=0.35~~,  \cr
p_{_F} = 0.49 \ GeV, \qquad \bar E_{B^*} = 5.65 \ GeV\qquad
&{\rm for}\; \alpha_s=0.24~~. }
\eqno(36)
$$

The calculated values of the $B$-meson mass, 5.42GeV ($\alpha_s = 0.35$) and
5.56GeV ($\alpha_s = 0.24$) are in reasonable agreement compared to the
experimental value of $m_B=5.28$GeV; the relative errors are 2.7\% and 5.3\%,
respectively.
But for Fermi motion parameter $p_{_F}$, the calculated values, 0.54GeV
($\alpha_s = 0.35$) and  0.49GeV
($\alpha_s = 0.24$), are somewhat larger than the value 0.3, widely used
in the experimental
analyses of energy spectrum of semileptonic $B$ meson decay.
The value $p_{_F} = 0.3$ corresponds to the $B$-meson radius
$R_B \sim 0.66$ fm,  and it
seems too large to us. 
On the other hand, the value $p_{_F} = 0.5$ corresponds to
$R_B \sim 0.39$ fm, which looks  in reasonable range.

The Fermi  motion parameter $p_{_F}$ is the most important
parameter of
ACCMM model, and the value $p_{_F} \sim 0.3$ GeV has been widely used in
experimental  analyses without clear theoretical or experimental evidence.
Therefore, it is strongly required to determine the value of $p_{_F}$ more
firmly when we  think of the importance of its role in experimental analyses.
We attempted to extract the possible
value for $p_{_F}$ theoretically in the relativistic quark model using with
quantum mechanical variational method. The better
determination of $p_{_F}$ is also interesting theoretically since it has
the physical correspondence related to the Fermi motion inside B-meson.
We obtained $p_{_F} \sim 0.5$ GeV, which is
somewhat larger than 0.3, and we also derived the ground state eigenvalue of
$E_B \simeq 5.5$ GeV, which is is reasonable agreement with $m_B=5.28$ GeV.

\bigskip
\centerline{\bf Acknowledgements}
\medskip

The work  was supported
in part by the Korean Science and Engineering  Foundation,
in part by Non-Direct-Research-Fund, Korea Research Foundation 1993,
in part by the Center for Theoretical Physics, Seoul National University,
in part by Yonsei University Faculty Research Grant,  and
in part by the Basic Science Research Institute Program, Ministry of Education,
  1994,  Project No. BSRI-94-2425.

\bigskip\bigskip\bigskip
\centerline{\bf Appendix}
\medskip

The integration constants $c_1$ and $c_2$ in (29) are given by the following
relations,
$$ \eqalignno{
I(0) &= -{1\over 2}c_2 = \int_0^\infty x^3e^{-x^2}dx = {1\over 2}~~, &(A-1) \cr
I''(s\approx 0) &= 2c_1 + c_2 (-{1\over 8}\ln s -{11\over 32}) \cr
&= -{1\over 4} \int_0^\infty x^2(x^2+s)^{-3/2}e^{-x^2}dx \quad {\rm at}
\quad s\approx 0 ~~. &(A-2)} 
 $$
Then, from (A-1), we get
$$
c_2~~=~~-1~~.
\eqno(A-3)
$$
The integral in (A-2) can be expanded as
$$ \eqalignno{
J(s=a^2) &=\int_0^\infty x^2(x^2+a^2)^{-3/2}e^{-x^2}dx \cr
&= \int_0^\infty x^2\big[(x+a)^2-2ax\big]^{-3/2}e^{-x^2}dx \cr
&= \int_0^\infty x^2(x+a)^{-3}\left[ 1-{2ax\over (x+a)^2} \right]^{-3/2}
e^{-x^2}dx \cr
&= \sum_{n=0}^\infty {(2n+1)!a^n \over 2^n(n!)^2}\int_0^\infty {x^{n+2}\over
(x+a)^{2n+3}}e^{-x^2}dx~~. &(A-4) } 
$$
Next the integral in (A-4) is obtained by 
$$
\int_0^\infty {x^{n+2}\over (x+a)^{2n+3}}e^{-x^2}dx = {1\over (2n+2)!}
\left({\partial\over\partial a}\right)^{2n+2}
\int_0^\infty {x^{n+2}\over x+a}e^{-x^2}dx~~.
\eqno{(A-5)}
$$
Again the integral in (A-5) is related to another integral, for a small value
of $a$,
$$
\int_0^\infty{x^{n+2}\over x+a}e^{-x^2}dx = \sum_{k=0}^{n+1}{(-a)^k \over
2} \left({n-k\over 2}\right)!+(-a)^{n+2}\int_0^\infty{e^{-x^2}\over x+a}dx~~.
\eqno(A-6)
$$
The integral in (A-6) can be expanded in a similar way as to obtain the series
(29) by making use of differential eqations. For a small value of $a$,
$$ 
\int_0^\infty {e^{-x^2}\over x+a} dx = -{1\over 2}e^{-a^2} (2\ln
a + \gamma + a^2 + {1\over 2} {a^4\over 2!} + \cdots  ) 
+ \sqrt \pi e^{-a^2} (a + {1\over 3}a^3 + {1\over 5}{a^5\over 2!} +\cdots )~~,
\eqno(A-7)
$$
where $\gamma \sim 0.5772$ is the Euler's constant.
In this way the constant $c_1$ is given by an infinite series,
$$
c_1 = -{3\over 64} + {\gamma \over 16} - {1\over 8}\sum_{n=1}^\infty
{1\over n2^n} \approx -0.0975 ~~.
\eqno(A-8)
$$

\vfill\eject

\centerline{\bf References}
\medskip

\item{[1]} For example, see H. Quinn,
in the proceedings of the third KEK topical conference on CP violation,
Nov. 1993, Tsukuba, to be published in Nucl. Phys. B (Proc. Suppl.) (1994);
L. Hall, $ibid.$

\item{[2]} CLEO Collab., R. Fulton et al., Phys. Rev. Lett. 64 (1990) 16.

\item{[3]} ARGUS Collab., H. Albrecht et al., Phys. Lett. B 234 (1990)
409; B 241 (1990) 278.

\item{[4]} V. Barger, C.S. Kim and R.J.N. Phillips, Phys. Lett. B 235
(1990) 187; B 251 (1990) 629; C.S. Kim, D. Hwang, P. Ko and W. Namgung,
in the proceedings of the third KEK topical conference on CP violation,
Nov. 1993, Tsukuba, to be published in Nucl. Phys. B (Proc. Suppl.) (1994);
C.S. Kim, P. Ko, Daesung Hwang and Wuk Namgung, SNUTP 94--49 (May 1994).

\item{[5]} G. Altarelli, N. Cabbibo, G. Corbo, L. Maiani and G.
Martinelli, Nucl. Phys. B 208 (1982) 365.

\item{[6]} I. I. Bigi, M. Shifman, N.G. Uraltsev and A. Vainshtein, Phys. Rev.
Lett.
71 (1993) 496.

\item{[7]} K. Hagiwara, A.D. Martin and A.W. Peacock, Z. Phys. C33 (1986) 135.

\end